\documentclass[journal]{IEEEtran}

\IEEEoverridecommandlockouts
\usepackage{amsmath,amssymb,amsfonts}
\usepackage{graphicx}
\usepackage{textcomp}
\usepackage{xcolor}
\usepackage{comment}
\usepackage{authblk}

\usepackage{multicol}
\usepackage[ruled,vlined,linesnumbered]{algorithm2e}
\usepackage{color}
\usepackage{graphicx}  
\usepackage{soul}
\usepackage{physics}
\def\vahid{\textcolor{blue}}

\usepackage{comment}
\usepackage[numbers,sort&compress,square]{natbib}
\usepackage{enumerate}

\usepackage[numbers,sort&compress,square]{natbib}
\usepackage{pdfpages}

\usepackage[normalem]{ulem}
\usepackage{float}

\usepackage{scalerel}
\usepackage{tikz}
\usetikzlibrary{svg.path}

\definecolor{orcidlogocol}{HTML}{A6CE39}
\tikzset{
  orcidlogo/.pic={
    \fill[orcidlogocol] svg{M256,128c0,70.7-57.3,128-128,128C57.3,256,0,198.7,0,128C0,57.3,57.3,0,128,0C198.7,0,256,57.3,256,128z};
    \fill[white] svg{M86.3,186.2H70.9V79.1h15.4v48.4V186.2z}
                 svg{M108.9,79.1h41.6c39.6,0,57,28.3,57,53.6c0,27.5-21.5,53.6-56.8,53.6h-41.8V79.1z M124.3,172.4h24.5c34.9,0,42.9-26.5,42.9-39.7c0-21.5-13.7-39.7-43.7-39.7h-23.7V172.4z}
                 svg{M88.7,56.8c0,5.5-4.5,10.1-10.1,10.1c-5.6,0-10.1-4.6-10.1-10.1c0-5.6,4.5-10.1,10.1-10.1C84.2,46.7,88.7,51.3,88.7,56.8z};
  }
}

\newcommand\orcidicon[1]{\href{https://orcid.org/#1}{\mbox{\scalerel*{
\begin{tikzpicture}[yscale=-1,transform shape]
\pic{orcidlogo};
\end{tikzpicture}
}{|}}}}

\def\BibTeX{{\rm B\kern-.05em{\sc i\kern-.025em b}\kern-.08em
    T\kern-.1667em\lower.7ex\hbox{E}\kern-.125emX}}
\begin{document}

\title{A Novel Optimal Modulation Strategy for Modular Multilevel Converter Based HVDC Systems}



\author{Saroj Khanal and Vahid R. Disfani \\
University of Tennessee at Chattanooga, Chattanooga, TN 37403 USA \\
emails: saroj-khanal@mocs.utc.edu, vahid-disfani@utc.edu 

\thanks{
\textcopyright 2019 IEEE. Personal use of this material is permitted. Permission from IEEE must be obtained for all other uses, in any current or future media, including reprinting/republishing this material for advertising or promotional purposes, creating new collective works, for resale or redistribution to servers or lists, or reuse of any copyrighted component of this work in other works. \\

Accepted and presented to the 2019 IEEE 2nd International Conference on Renewable Energy and Power Engineering.}}



\maketitle

\begin{abstract}
Unlike conventional converters, modular multilevel converter (MMC) has a higher switching frequency -- which has direct implication on important parameters like converter loss and reliability -- mainly due to increased number of switching components. However, conventional switching techniques, where submodule sorting is just based on capacitor voltage balancing, are not able to achieve switching frequency reduction objective. A novel modulation algorithm for modular multilevel converters (MMCs) is proposed in this paper to reduce the switching frequency of MMC operation by defining a constrained multi-objective optimization model. The optimized switching algorithm incorporates all control objectives required for the proper operation of MMC and adds new constraints to limit the number of submodule switching events at each time step. Variation of severity of the constraints leads to a desired level of controllability in MMC switching algorithm to trade-off between capacitor voltage regulation and switching frequency reduction. Finally, performance of the proposed algorithm is validated against a seven-level back-to-back MMC-HVDC system under various operating conditions.

\end{abstract}

\begin{IEEEkeywords}
Capacitor voltage balancing, high voltage direct current (HVDC), model predictive control (MPC), optimal switching, sorting algorithm.
\end{IEEEkeywords}

\section{Introduction}
Modular multilevel converter (MMC) has been well known as a preferred choice among converter topologies for medium/high-power applications, mainly thanks to its salient features such as modularity, scalability, high efficiency, high reliability, and superior harmonic performance \cite{qin2012predictive, debnath2015operation}. Among various well-known applications, it has particularly become the worldwide standard for voltage-sourced converter high-voltage direct current (VSC-HVDC) transmission systems \cite{marquardt2018modular}.

MMC should actively meet multiple control objectives for its proper operation. These include objectives of fulfilling submodule (SM) capacitor voltage balancing, output AC current tracking, and circulating current mitigation/elimination. Switching frequency of MMC is significantly higher than those of conventional and other multilevel converter topologies, mainly due to the increased number of switching devices. As switching frequency has direct implication on converter losses, switching design has an important role in shaping MMC's future \cite{debnath2015operation}. Among modulation methods, predictive methods (also called model predictive control (MPC) based methods) have lately attracted a significant amount of attention among academia and industry, mainly due to their fast dynamic response and ability to meet the multiple control objectives with an increased flexibility \cite{qin2012predictive,disfani2015fast, gong2016design, rodriguez2012predictive, ma2014one, dekka2018model, moon2014model}. However, the predictive methods suffer from two major issues: (i) computational complexity and (ii) higher switching-frequency operation. The first makes these methods impractical, while the latter makes them inappropriate for the use in high-power applications. There has been a significant efforts toward implementing computationally efficient/fast MPC methods \cite{ma2014one, disfani2015fast, gong2016design, liu2015grouping,zhang2016voltage,huang2017priority}; however, little attention has been given in comprehensively addressing reduced switching-frequency switching algorithms.

Several modulation strategies have been reported in the literature to reduce switching frequency \cite{tu2011reduced,qin2013reduced, dekka2017integrated}. A general framework, including slow-rate, hybrid, and fundamental voltage balancing strategies, to reduce switching frequency has been introduced in \cite{qin2013reduced}. The slow-rate method, however, cannot keep track of SM capacitor voltages. Similarly, an integrated-MPC method leveraging discontinuous modulation approach is introduced in \cite{dekka2017integrated}. A common main issue with these methods is the unnecessary high-switching transition of SMs in every control cycle as they fail to take into account the previous switching statuses. A modified phase-shifted carrier-based PWM (PSC-PWM) together with a reduced switching-frequency algorithm is proposed in \cite{tu2011reduced}, which limits bypassing/insertion of the SMs every control cycle allowing turned-off SMs for next cycle to meet voltage requirement. However, strategies like this -- which are among few that consider the previous switching statuses -- are not scalable and do not provide controllability to tune the priority of capacitor voltage balancing and switching frequency reduction against one another.


In this paper, a novel constrained, multi-objective optimization model is proposed that simultaneously regulate SM capacitor voltages and minimize switching frequency. This algorithm takes the previous statuses of SMs into account while sorting the SMs for selection process. It first sorts SMs based on their capacitor voltage values, then applies the maximum switching frequency constraint to avoid extensive number of switching events. Finally, the paper also analyzes the effects of this constraint on SM capacitor voltages. 


\section{Mathematical Models for Modular Multilevel Converters} \label{sec:Math_model_MMC}
\begin{figure}
\centering
\includegraphics[width=0.45\textwidth]{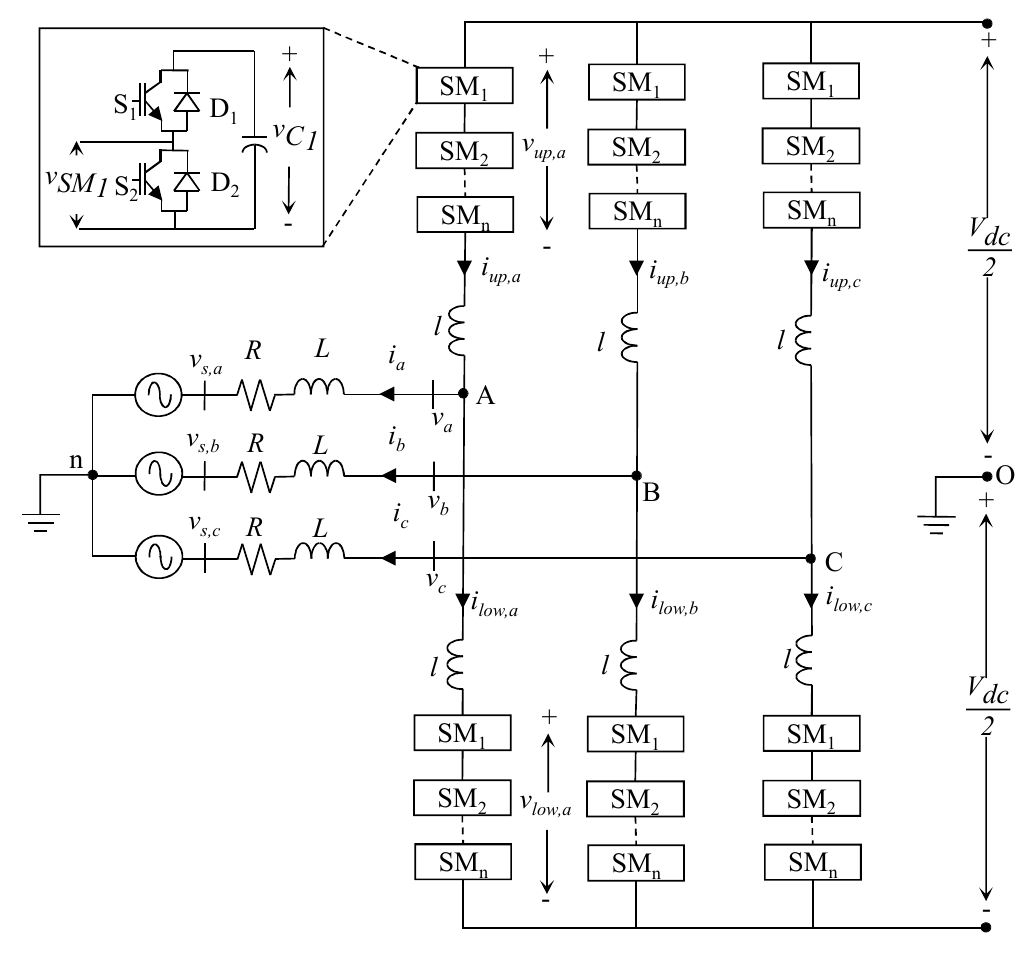}
\caption{Circuit diagram of a modular multilevel converter.}
\label{MMC}    
\end{figure}

A schematic diagram of an $(n+1)$-level, three-phase MMC based on half-bridge SMs is depicted in Fig.~\ref{MMC}. Each phase (leg) consists of two arms with $n$ submodules (SMs) on each of them. MMC considered in this paper is made up of half-bridge SMs and three arm inductors ($l$) to reduce surge and fault currents. The MMC is connected to a three-phase AC system through a transformer, which is modeled as an equivalent series resistive-inductive ($R-L$) impedance.
\subsection{Discrete-time Model of MMC} \label{MMC-discrete}
In this paper, the discrete model of MMC that is derived in the authors' previous paper \cite{disfani2015fast} is used to develop the corresponding algorithms. In this model, the next step value for the AC-side current for a sufficiently small sampling time step $T_s$ is derived as:
\begin{align}
&\begin{matrix}i(t+T_s)=\frac{1}{K'}{\left(  {\frac {v_{low}(t+T_s)-v_{up}(t+T_s)}{2} -v_s(t+T_s)+\frac{L'}{T_s}i(t)}\right)}\end{matrix}
\label{Idis}
\end{align}

\noindent where $L'=L+l/2$, $K'=R+L'/T_s$, and the time indices $(t)$ and $(t+T_s)$ are respectively used to denote measured values at the current time step and the anticipated values for the next time step. Assuming that sampling frequency is sufficiently higher than the grid frequency, we can replace the predicted value of grid voltage $v_s(t+T_s)$ by its measured value $v_s(t)$. Denoting the status of $j$-th SM by $u_j(t+T_s)$, anticipated capacitor voltage of individual SMs on upper-level and lower-level arms are calculated as:
\begin{align}
&v_{Cj}(t+T_s)=v_{Cj}(t)+\left(\frac{T_s i_{up}(t)}{C}\right)u_j(t+T_s) && \forall_{j\in [1,n]}\label{VCupdis} \\
&v_{Cj}(t+T_s)=v_{Cj}(t)+\left(\frac{T_s i_{low}(t)}{C}\right)u_j(t+T_s) &&\forall_{j\in [n+1,2n]}. \label{VClowdis}
\end{align}

Also, anticipated voltages across upper-level and lower-level arms as well as circulating current are equal to:
\begin{align}
&v_{up}(t+T_s)=\sum_{j=1}^{n} v_{Cj}(t+T_s)u_j(t+T_s)  \label{Vupdis} \\
&v_{low}(t+T_s)=\sum_{j=n+1}^{2n} v_{Cj}(t+T_s)u_j(t+T_s) \label{Vlowdis} \\
&i_z(t+T_s)=\frac{T_s}{2l}\left(V_{dc}-v_{low}(t+T_s)-v_{up}(t+T_s)\right)+i_z(t).
\label{Izdis}
\end{align}

\section{Constrained Switching-Frequency Model Predictive Control for MMC} \label{sec:algorithms}

\subsection{Optimization Model}
The following four objectives are considered for the modulation and control design of MMC \cite{qin2012predictive, disfani2015fast}: 
\begin{enumerate}[i.]
\item to regulate all SM capacitor voltages to their nominal values ($V_{dc}/n$),
\item to track the AC-side current ($i$) of all phases satisfactorily around their reference values ($i_{ref}$),
\item to mitigate the circulating current ($i_z$) flowing among the phase legs, and
\item to limit switching frequency less than a desired frequency.
\end{enumerate}

The first three objectives are common among the predictive methods and have been addressed in authors' previous works \cite{disfani2015fast,ma2014one}. Development of the last objective and application of it as a new constraint to the optimization problem is the main contribution of this paper.


Assuming that exact AC current waveform tracking $i(t+T_s)=i_{ref}$ and exact circulating current suppression $i_z(t+T_s)=0$ can be achieved, the target values of upper-level and lower-level voltages of MMC are calculated as
\begin{align}
&v_{up}^*=\left(\frac{V_{dc}}{2}+\frac{l}{T_s}i_z(t)\right)-
\left(K'i_{ref}+v_s(t)-\frac{L'}{T_s}i(t)\right)
\label{Vup*}\\
&v_{low}^*=\left(\frac{V_{dc}}{2}+\frac{l}{T_s}i_z(t)\right)+
\left(K'i_{ref}+v_s(t)-\frac{L'}{T_s}i(t)\right)
\label{Vlow*}
\end{align}
where $(\cdot)^*(t+T_s)$ denotes the ideal anticipated value of the corresponding variable. Deviation of actual AC current waveform and circulating current from their target values are equal to
\begin{align}
&\Delta i= \frac{1}{2K'}\left(\Delta v_{low}(t+T_s)-\Delta v_{up}(t+T_s)\right)
\label{Idis_error}\\
&i_z(t+T_s)=\frac{T_s}{2l}\left(\Delta v_{low}(t+T_s)+\Delta v_{up}(t+T_s)\right)
\label{Izdis_error}
\end{align}
where $\Delta i=i-i_{ref}(t+T_s)$, $\Delta v_{low}=v_{low}^*-v_{low}$, and $\Delta v_{up}=v_{up}^*-v_{up}$.

Using weighted sum method to combine the AC current waveform tracking and circulating current mitigation objectives with weights $w$ and $w_z$ respectively, a multi-objective optimization problem with the formulation below is developed to describe the proposed switching algorithm:

\begin{align}
&\min_U&& {\sum_{j=1}^{2n}\abs{v_{C_j}(t+T_s)-v_{C_j}(t)}}  \label{sorting_volt_obj}\\
&\min_U&&f=\left\{\begin{matrix}\frac{w}{2K'}\left|\Delta v_{low}(t+T_s)-\Delta v_{up}(t+T_s)\right|+\\
\\
\frac{w_z T_s}{2l}\left|\Delta v_{low}(t+T_s)+\Delta v_{up}(t+T_s)\right|\end{matrix}\right\}\label{selection_obj}\\
&\text{subject to:} &&~~\eqref{Idis}-\eqref{Izdis}\nonumber\\
&&& ~~\sum_{j=1}^{n} \abs{u_j(t+T_s)-u_j(t)}\leq \overline{N_{sw}}\label{constraint_upper_sw}\\
&&& \sum_{j=n+1}^{2n} \abs{u_j(t+T_s)-u_j(t)}\leq \overline{N_{sw}} \label{constraint_lower_sw}\\
&&& ~~U=[u_1,u_2,...,u_{2n}] : u_j \in \{0,1\}~~~~\forall_{j\in[1,2n]}.\label{constraint_status_binary}
\end{align}
In this model, \eqref{sorting_volt_obj} addresses SM capacitor voltage regulation, \eqref{constraint_upper_sw} and \eqref{constraint_lower_sw} limit the number of switching events on each arm to be less than the desired value of $\overline{N_{sw}}$, and \eqref{selection_obj} fulfills AC current waveform tracking and circulating current mitigation objectives. 

\subsection{Solution Algorithm}
\begin{figure}[htb]
    \centering
    \includegraphics[width=0.45\textwidth]{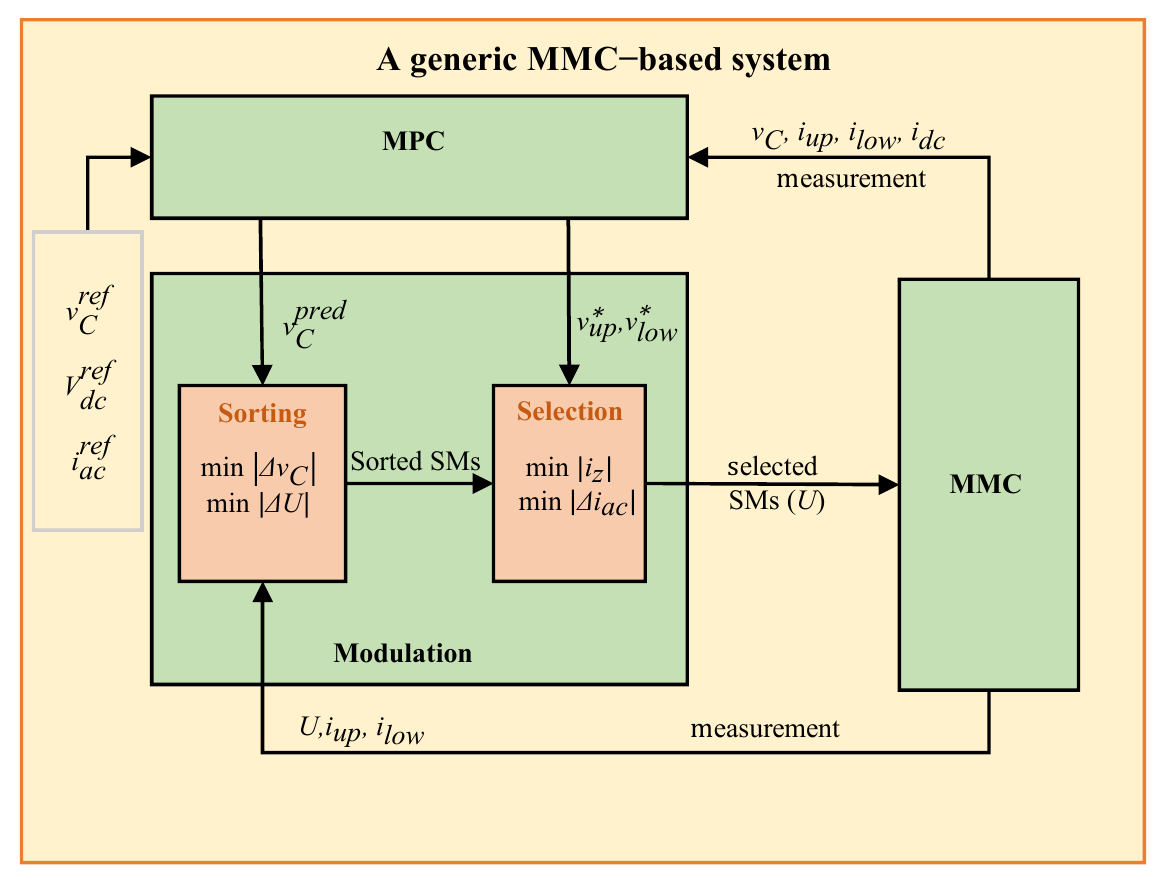}
    \caption{\label{fig:MMCSys} Block diagram of operation of switching algorithms for MMC-based systems.}
\end{figure}

Similar to \cite{disfani2015fast}, two steps of SM sorting and SM selection are defined to solve the multi-objective optimization problem as shown in Fig. \ref{fig:MMCSys}. 

\subsubsection{Step 1 -- Submodule Cascaded Sorting}
The objective function \eqref{sorting_volt_obj} and the constraints \eqref{constraint_upper_sw} and \eqref{constraint_lower_sw} are targeted in this step by sorting SMs effectively such that the highest priority is given to SMs contributing the most in SM voltage balancing and switching frequency reduction without violating the maximum switching event and maximum SM voltage deviation constraints. SMs of both upper and lower arms are first sorted based on their anticipated capacitor voltages to fulfill the voltage balancing objective. The direction of $i_{up}$ determines whether the capacitor voltages of upper-level submodules tend to increase or decrease at the next time step. Thus, the algorithm sorts SMs based on their capacitor voltages in ascending order if $i_{up}\ge 0$ and in descending order otherwise. 
%
%

To address the constraints on number of switching events, any solution that violates these constraints must be eliminated from the feasible set. However, to avoid infeasibility, this paper relaxes these constraints by applying the Lagrange multipliers and transferring the constraints to the objective function with a priority higher than that of the voltage balancing objective function. After SMs are sorted based on the objective functions, accumulated sum of number of switching events on each arm is calculated for each possible solution $m$ and is called $N_{sw_m}$ where solution $m$ corresponds to selection of first $m$ sorted SMs to be switched on at the next time step. That is, $N_{sw_m}=\sum_{j=1}^{m} \abs{u_j(t+T_s)-u_j(t)}$ for upper arm and $N_{sw_m}=\sum_{j=n+1}^{n+m} \abs{u_j(t+T_s)-u_j(t)}$ for lower arm if $u_j(t+T_s)$ is the status of SM $j$ in solution $m$. SMs are then sorted in an ascending order based on whether or not their $N_{sw_m}$ violate the maximum switching constraint, i.e.,
\begin{align}
\mu_{sw_m}=\max\{0,N_{sw_m}-\overline{N_{sw}}\}. \label{Nsw}
\end{align}

This procedure is detailed in \textbf{Algorithm~\ref{V1-FC}} and is called F1-VC hereafter. Conventional sorting algorithm focused on voltage balancing is called V1-F2 in this paper and is used as the benchmark algorithm in the case study.

\begin{algorithm}
\DontPrintSemicolon
\caption{Constrained-Switching-Frequency Voltage-Balancing Algorithm (V1-FC)}
\label{V1-FC}
\For{all phases $a,b,c$}{
    Collect measurements of capacitor voltages, arm currents, and current switching status ($u_j^{curr}$)\;
    Sort SMs based on $u_j^{curr}$ in descending order\;
    Calculate anticipated SM voltages $v_{C_j}$\;
    \For{$k\in\{up,low\}$}{
    \eIf{$i_{k}\ge 0$}
    {Sort the sorted SMs based on $v_{C_j}$ in ascending order\;}
    {Sort the sorted SMs based on $v_{C_j}$ in descending order\;}}
    Calculate accumulated sum of switching events up to each sorted SM ($N_{sw_m}$) and $\mu_{sw_m}$ from \eqref{Nsw}\;
    Sort the sorted SMs based on $\mu_{sw_m}$ in ascending order\;}
Proceed to \textbf{Step 2}.
\end{algorithm}

\subsubsection{Step 2 -- Submodule Selection}
The SM selection algorithm introduced in \cite{disfani2015fast} is employed in this paper. Let us assume that the vectors $V_{C_{up}}^{sort}=[V_{C_1}^{sort},...,V_{C_n}^{sort}]$ and $V_{C_{low}}^{sort}=[V_{C_{n+1}}^{sort},...,V_{C_{2n}}^{sort}]$ denote corresponding SM voltages on upper and lower arms, after being sorted in Step 1. In this step, the algorithm calculates the cumulative sum vectors of the components of $V_{C_{up}}^{sort}$ and $V_{C_{low}}^{sort}$ to establish the sets $V_{C_{up}}^{sum}$ and $V_{C_{low}}^{sum}$ as:
\begin{align}
&V_{C_{up}}^{sum}=\{\alpha_k:k=0,1,...,n\} \label{sum_up}\\
&V_{C_{low}}^{sum}=\{\beta_k: k=0,1,...,n\}  \label{sum_low}
\end{align}
where
\begin{align}
&\alpha_0=\beta_0=0\nonumber\\
&\alpha_k=\Sigma_{i=1}^{k}V_{C_i}^{sort}&\forall_{k\in [1,n]}\nonumber\\
&\beta_k=\Sigma_{i=n+1}^{n+k}V_{C_i}^{sort}&\forall_{k\in [1,n]}.\nonumber
\end{align}
Then, the switching algorithm determines which combination of $(\alpha,\beta)$ results in minimum value of the objective function \eqref{selection_obj}. Reference \cite{disfani2015fast} provides a mathematical proof which states that if $v_{up}^*\in [\alpha_i,\alpha_{i+1})$ and $v_{low}^*\in [\beta_j,\beta_{j+1})$, the optimal solution belongs to the set $\{(\alpha_i,\beta_j),(\alpha_{i+1},\beta_j), (\alpha_i,\beta_{j+1}),(\alpha_{i+1},\beta_{j+1})\}$. It greatly improves the efficiency of the switching algorithm since it suffices to check the objective function for just 4 possible solutions to find the best option, instead of checking all $n^2$ feasible combinations of $\alpha$ and $\beta$.
\section{Case Study} \label{sec:case_study}
\subsection{Test System}
\begin{figure}[htb]
    \centering
    \includegraphics[width=0.45\textwidth]{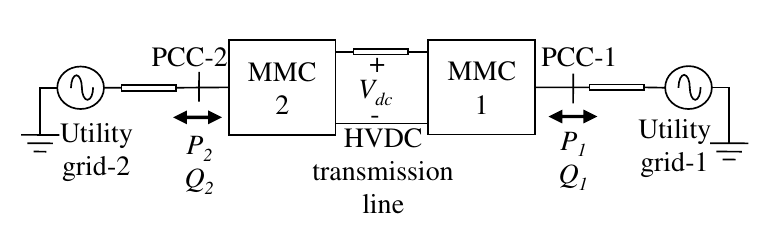}
    \caption{Schematic representation of an MMC-based back-to-back HVDC System.}
    \label{fig:HVDC_SLD}
\end{figure}

The proposed algorithm is tested against an HVDC system similar to the one illustrated in Fig.~\ref{fig:HVDC_SLD}. In this HVDC system, MMC2 acts as a controlled DC power source. MMC1 is the test MMC for both modulation and control systems. The system parameters are provided in the Table~\ref{tab:sys_param}. The performance of the proposed algorithm is benchmarked against that of the conventional voltage balancing algorithm, called as V1-F2 hereafter.

\begin{table}[ht]
    \centering
    \caption{Case study parameters}
    \begin{tabular}{|l|l|}
    \hline
        Quantity & Value \\
        \hline
        Number of submodules per arm & 6 \\ 
        MMC nominal power & 50 MVA \\
        Nominal DC voltage $(V_{dc})$ & 60 kV \\
        Submodule capacitor ($C_{sm}$) & 2.5 mF\\
        Active Power Transferred ($P_{1}$) & 13.18 MW \\
        $R$ & 0.03 $\Omega$\\
        $L$ & 5 mH\\
        $l$ & 3 mH\\
        Sampling period ($T_s$) &25 $\mu$s\\
        HVDC line length &5 km\\
        HVDC link capacitor &16 $\mu$F/km\\
        HVDC line inductance &50 $\mu$H/km\\
        \hline
    \end{tabular}
    \label{tab:sys_param}
\end{table}

In this case study, V1-FC algorithm is tested where the maximum number of switching events at each time step is constrained at an integer number ($\overline{N_{sw}}$) between zero and number of submodules on each arm (6 in this test case) as given by \eqref{n_sw}. 

\begin{align}
\overline{N_{sw}}=\left\{\begin{matrix}
 6& 1~\mathrm{s}<t\leq 1.2~\mathrm{s}\\ 
 0& 1.2~\mathrm{s}<t\leq 1.4~\mathrm{s}\\ 
 1& 1.4~\mathrm{s}<t\leq 1.6~\mathrm{s}\\ 
 2& 1.6~\mathrm{s}<t\leq 1.8~\mathrm{s}\\ 
 3& 1.8~\mathrm{s}<t\leq 2~\mathrm{s}\\ 
 4& 2~\mathrm{s}<t\leq 2.2~\mathrm{s}\\ 
 5& 2.2~\mathrm{s}<t\leq 2.4~\mathrm{s}\\ 
 6& 2.4~\mathrm{s}<t\leq 2.6~\mathrm{s}.
\end{matrix} \right.
\label{n_sw}
\end{align}

\subsection{Results and Discussions}

\begin{figure}[htb]
    \centering
    \includegraphics[width=0.489\textwidth]{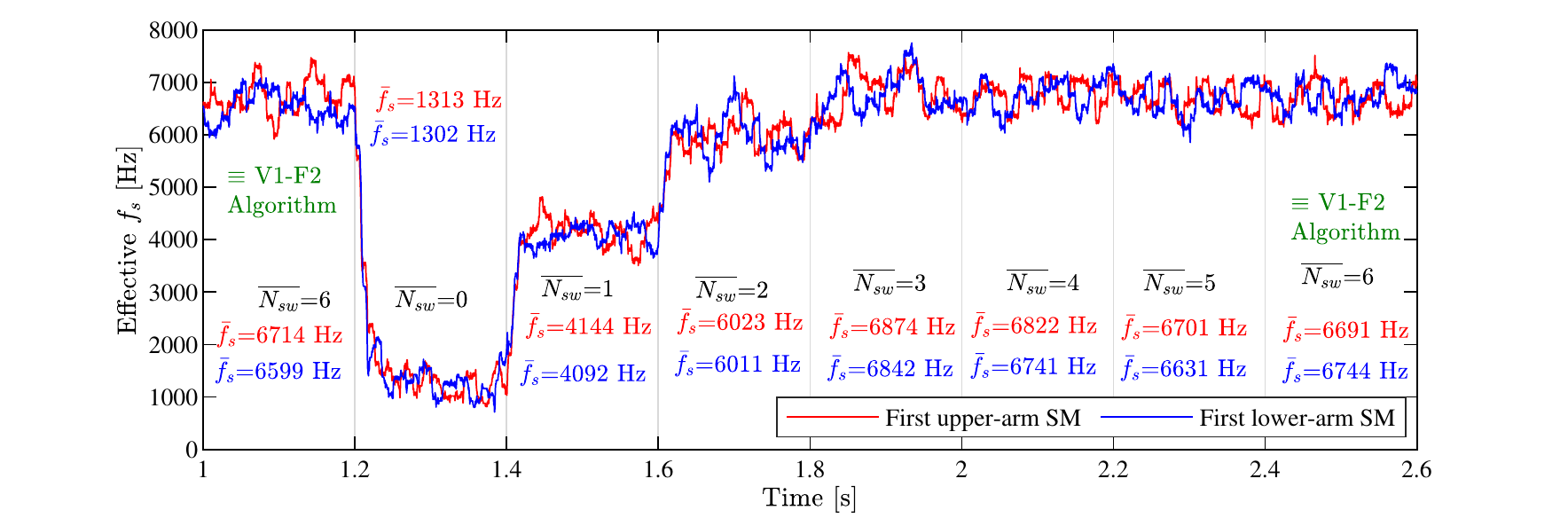}
    \caption{Effective switching frequencies observed at phase-leg A SMs while applying V1-FC algorithm with varying $\overline{N_{sw}}$.}
    \label{fig:f_s}
\end{figure}

Fig.~\ref{fig:f_s} shows the effective switching frequencies observed in the first submodules of upper- and lower-arm of phase-leg A with the varying $\overline{N_{sw}}$. Average steady-state effective switching frequencies are calculated and shown for both first upper- and lower-arm submodules of phase-leg A. It clearly shows 80\% reduction in effective switching frequencies, with $\overline{N_{sw}}=0$. Reductions in effective switching frequency for different values of $\overline{N_{sw}}=1$ and $\overline{N_{sw}}=2$ are equal to 38\% and 10\%, respectively. For $\overline{N_{sw}}\in\{3,4,5\}$, the reduction in effective switching frequency is up to 3\%.

\begin{figure}[htb]
    \centering
    \includegraphics[width=.47\textwidth]{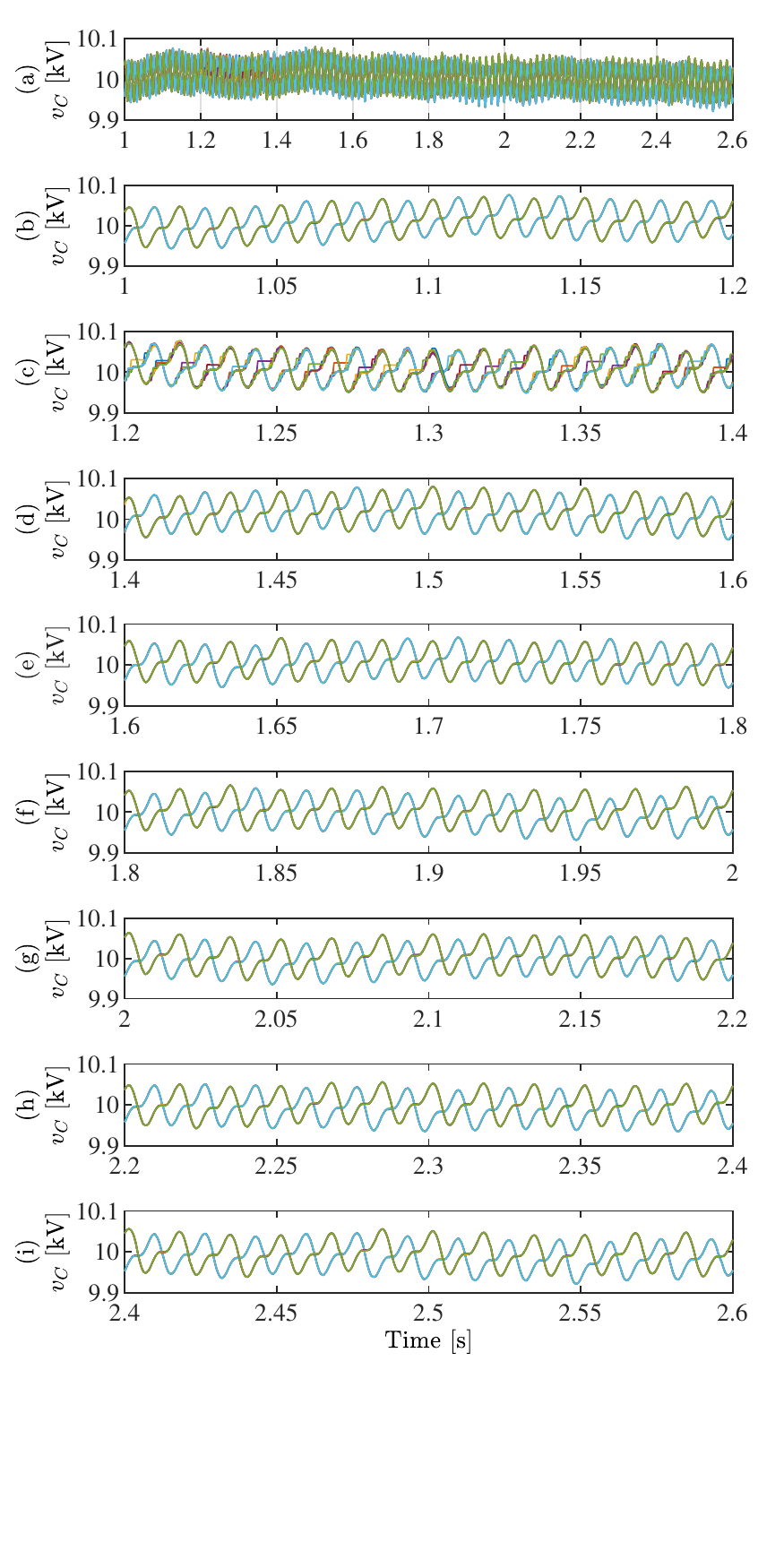}
    \vspace{-0.9in}
    \caption{SM capacitor voltages of phase-leg A with V1-FC algorithm with varying values of $\overline{N_{sw}}$: (a) $\overline{N_{sw}}\in\{0,1,\cdots,6\}$ as described by \eqref{n_sw} (b)(i) $\overline{N_{sw}}=6$, (c) $\overline{N_{sw}}=0$, (d) $\overline{N_{sw}}=1$, (e) $\overline{N_{sw}}=2$, (f) $\overline{N_{sw}}=3$, (g) $\overline{N_{sw}}=4$, and (h) $\overline{N_{sw}}=5$.}
    \label{fig:v_C}
\end{figure}

Fig.~\ref{fig:v_C} shows the capacitor voltages of both upper and lower arms of the phase-leg A of MMC1. V1-FC algorithm with no restriction on switching transition ($\overline{N_{sw}}=6$) equivalently represents the conventional V1-F2 algorithm. In this condition, the capacitor voltages of all submodules on upper (or lower) arm change altogether as a result of the voltage balancing strategy employed. Between $t=1.2~\mathrm{s}$ and $t=1.4~\mathrm{s}$ where $\overline{N_{sw}}=0$, there are some differences among and distortions on individual capacitor voltages of upper (or lower) arm SMs. This is because the first priority in this switching algorithm is given to reduce switching frequency, and thence voltage balancing objective is compromised. When $\overline{N_{sw}}=1$ during  $t=1.4~\mathrm{s}$ and $t=1.6~\mathrm{s}$, the effect on SM capacitor voltages is significantly reduced due to the relaxation of constraints \eqref{constraint_upper_sw} and \eqref{constraint_lower_sw}. 
In this way, the algorithm provides full controllability over the major trade-off of capacitor voltage. In addition, the voltage ripples of SM capacitors are all the same (around 1.2\%) and are not compromised by the switching algorithm employed. 

\begin{figure}[htb]
    \centering
    \includegraphics[width=0.45\textwidth]{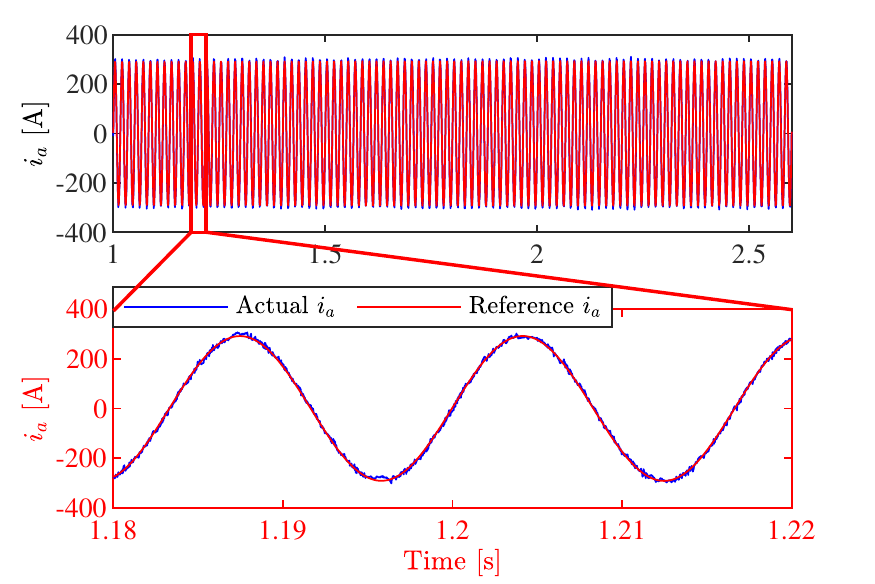}
    \caption{AC output current through terminal A of MMC1 ($i_a$) while applying V1-FC algorithm with varying $\overline{N_{sw}}$.} 
    \label{fig:i_a}
\end{figure}

Fig.~\ref{fig:i_a} shows the waveforms of reference AC current and actual output AC current at the terminal A of MMC1. It illustrates how AC reference current tracking is achieved to fulfill the second objective for different values of $\overline{N_{sw}}$. 

\begin{figure}[htb]
    \centering
    \includegraphics[width=0.45\textwidth]{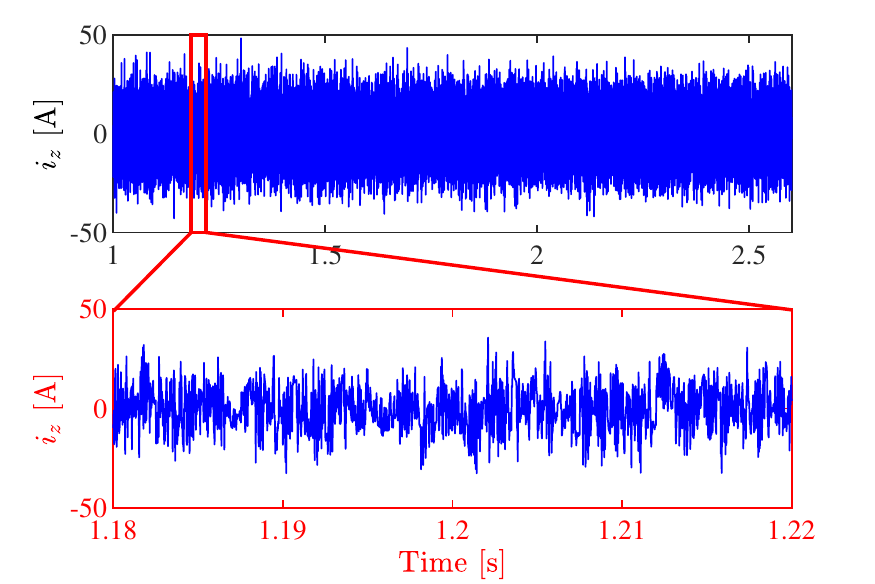}
    \caption{Circulating current through phase-leg A of MMC1 ($i_z$) while applying V1-FC algorithm with varying $\overline{N_{sw}}$.}
    \label{fig:i_z}
\end{figure}

Fig.~\ref{fig:i_z} illustrates the circulating current ($i_z$) flowing through terminal A of MMC1 during $t\in(1~\mathrm{s},2.6~\mathrm{s}]$. The results show that the circulating current is successfully controlled around zero Ampere and its maximum deviation is just 10\% of the magnitude of AC output current ($i_a$) of MMC. These results verify that the third objective is fulfilled for different values of $\overline{N_{sw}}$.

\section{Conclusion} \label{sec:conclusion}
A novel optimized-switching algorithm for model predictive control (MPC) of modular multilevel converter (MMC) is proposed to reduce switching frequency by applying constraints on the number of switching events at each time step. 
The algorithm is tested against a three-phase MMC based back-to-back HVDC system in MATLAB/Simulink to demonstrate its effectiveness under various switching constraints. 


\bibliographystyle{IEEEtran}
{\footnotesize \bibliography{IEEEabrv,MMC}}

\end{document}